\documentclass[a4paper,oneside,article]{article}
\usepackage[utf8]{inputenc}
\usepackage[english]{babel}
\usepackage{multirow}
\usepackage{color}
\usepackage{bm, bbm}
\usepackage[backend=bibtex,
style=authoryear,
citestyle=authoryear-comp,
natbib=true,
sorting=nyvt,
doi=false,isbn=false,url=false]{biblatex}
\usepackage{float}
\usepackage{mathtools}
\usepackage{makecell}
\usepackage{arydshln}
\usepackage[ruled,vlined]{algorithm2e}
\usepackage[top=1in, bottom=1.25in, left=1in, right=1in]{geometry}

\usepackage{xr-hyper}
\usepackage{authblk}

\RequirePackage{amsthm,amsmath,amsfonts,amssymb}
\RequirePackage[colorlinks,citecolor=blue,urlcolor=blue]{hyperref}
\RequirePackage{graphicx}
\newcommand\citesq[1]{[\cite{#1}]}

\makeatletter

\begin{document}

\author[1,2]{Ida Egendal}
\author[1,2]{Rasmus Froberg Brøndum}
\author[3]{Marta Pelizzola}
\author[3]{Asger Hobolth}
\author[1,2]{Martin Bøgsted}

\affil[1]{Center for Clinical Data Science, Aalborg University and Aalborg University Hospital, Aalborg, Denmark}
\affil[2]{Clinical Cancer Research Center, Aalborg University Hospital, Aalborg, Denmark}
\affil[3]{Department of Mathematics, Aarhus University, Aarhus, Denmark}

\date{}

\title{On the Relation Between Autoencoders and Non-negative Matrix Factorization, and Their Application for Mutational Signature Extraction}
\maketitle

\begin{abstract}
The aim of this study is to provide a foundation to understand the relationship between non-negative matrix factorization (NMF) and non-negative autoencoders enabling proper interpretation and understanding of autoencoder-based alternatives to NMF. Since its introduction, NMF has been a popular tool for extracting interpretable, low-dimensional representations of high-dimensional data. However, recently, several studies have proposed to replace NMF with autoencoders. This increasing popularity of autoencoders warrants an investigation on whether this replacement is in general valid and reasonable. Moreover, the exact relationship between non-negative autoencoders and NMF has not been thoroughly explored. Thus, a main aim of this study is to investigate in detail the relationship between non-negative autoencoders and NMF. We find that the connection between the two models can be established through convex NMF, which is a restricted case of NMF. In particular, convex NMF is a special case of an autoencoder. The performance of NMF and autoencoders is compared within the context of extraction of mutational signatures from cancer genomics data. We find that the reconstructions based on NMF are more accurate compared to autoencoders, while the signatures extracted using both methods show comparable consistencies and values when externally validated. These findings suggest that the non-negative autoencoders investigated in this article do not provide an improvement of NMF in the field of mutational signature extraction. 
\end{abstract}

\section{Introduction} 
Non-negative matrix factorization (NMF) is a popular tool for unsupervised learning \citesq{lee1999}. NMF factorizes a non-negative data matrix into a product of two non-negative matrices of lower dimension: a basis matrix consisting of basis vectors and a weight matrix consisting of the basis vector's weights for each observation in the data matrix. NMF has gained a strong footing in different scientific fields due to its high interpretability \citesq{alexandrov2013, hyperspec, AE_MU}. Specifically, NMF has proven to be a useful tool to derive mutational signatures from cancer genomics data. 

In mutational signature analysis, it is typically assumed that all mutations in a cancer genome are caused by mutagenic processes that leave a characteristic pattern of mutations in the genome. These patterns are denoted mutational signatures. Several signatures have been identified and linked to different mutagenic processes such as ultraviolet light exposure and tobacco smoking \citesq{etiologi}. \citet{alexandrov2013} proposed using NMF on mutational count data from cancer genomes to decipher the mutational signatures of the processes the patients have been exposed to throughout the development of the disease. NMF has since then been the dominating model for mutational signature extraction \citesq{alexandrov2020, MutationalPatterns, SigProfiler}. When extracting mutational signatures with NMF, the data matrix consisting of a number of patients' mutational profiles is decomposed into a matrix representing the signatures of mutagenic processes (basis vectors) and an exposure matrix dictating the number of mutations that can be attributed to each specific process in the mutational profiles of each patient (weight matrix).

Recently, several studies have proposed substituting NMF with non-negative autoencoders which are increasingly popular for dimensionality reduction \citesq{Hosseini, AE_MU, Khatib, NAE, LEMME2012}. This is also the case in mutational signature extraction \citesq{pei, MUSE-XAE}. \citet{pei} suggested using a sparse autoencoder to identify mutational signatures from cancer genomics data and generated estimates that were not only in concordance with existing literature but also correlated with observed exogenous exposures in a meaningful way. \citet{MUSE-XAE} suggested a hybrid architecture with a deep encoding and shallow decoding to relax the assumption of linearity NMF imposes on mutational signature extraction. However, they did not compare the results and performance to NMF. This trend of using non-negative autoencoders as an alternative to NMF with promising results prompts the question: What is the mathematical relation between autoencoders and NMF? And how do they compare in applications?

The aim of this study is to compare the performance of shallow, non-negative autoencoders to NMF in the field of mutational signatures. In particular, we compare the in- and out-of-sample reconstruction error, as well as the consistency of the extracted signatures from the tumor-normal whole genome sequences of 713 ovary, 311 prostate, and 523 uterus tumors from the Genomics England 100,000 Genomes Project (GEL) \citesq{GEL, GEL2}. Moreover, we show theoretically that shallow, non-negative autoencoders and NMF is a special case of NMF where the basis vectors are restricted to be convex combinations of columns in the data matrix \citesq{cvxnmf}. Based on this, we deduce how it impacts interpretation of the estimates generated by non-negative autoencoders in general and especially within the field of mutational signatures.

In Section \ref{sec:methods}, we characterize the mathematical relationship between autoencoders and NMF and introduce the framework for comparing the two models. In Section \ref{sec:res}, we compare the performance of NMF and autoencoders and demonstrate the mathematical equivalence between convex NMF and non-negative, shallow autoencoders. Lastly, we discuss and conclude on the results in Sections \ref{sec:disc} and \ref{sec:conc}.

\section{Methods}\label{sec:methods}
\begin{figure}[!t]
    \centering
    \includegraphics[width=0.65\linewidth]{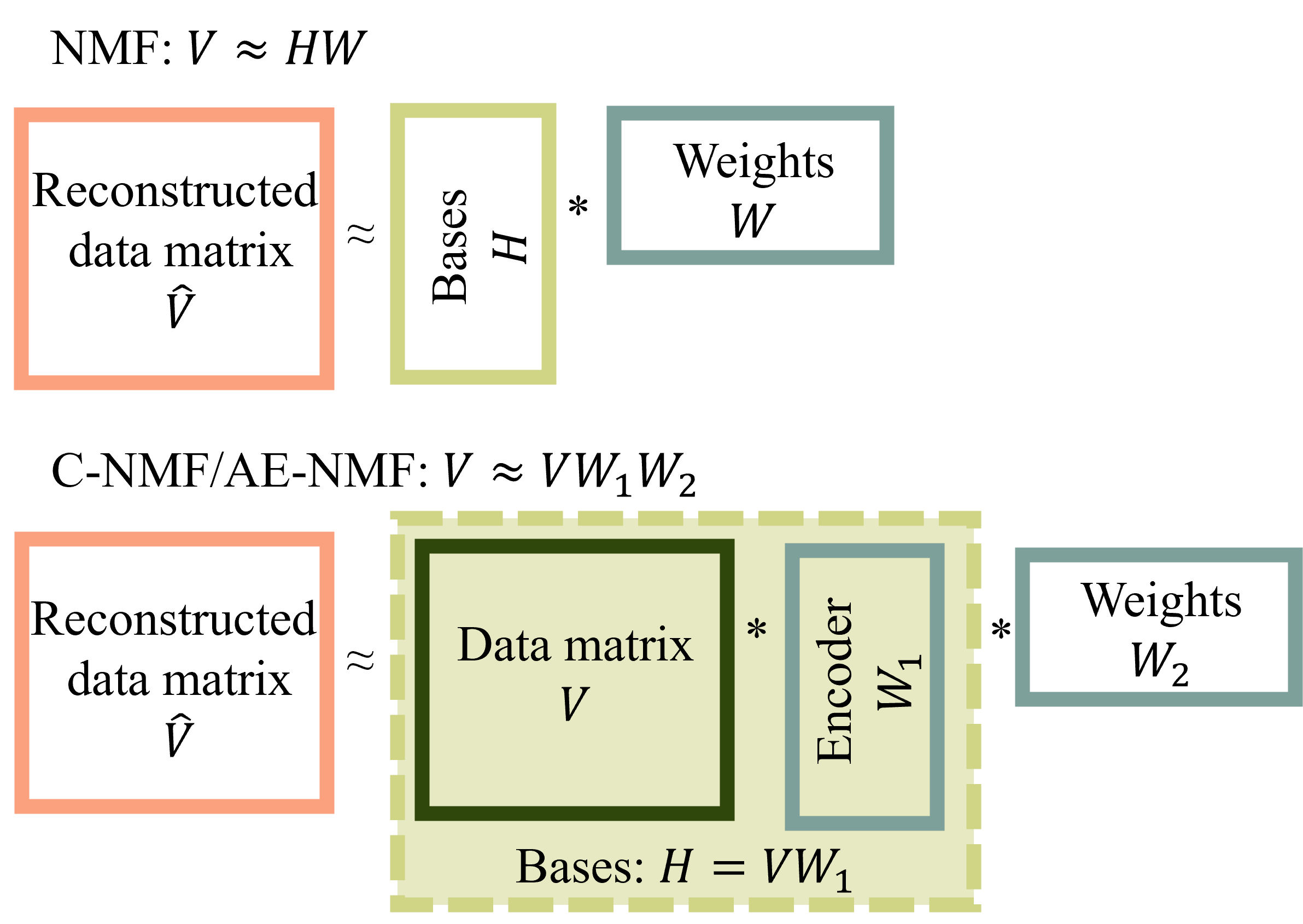}
    \caption{Schematic representation of the composition of basis vectors and weights in NMF (top) and C-NMF and AE-NMF (bottom).}
    \label{fig:AE_sceheme}
\end{figure}

Consider a non-negative data matrix $\bm{V}\in\mathbb{R}_+^{M\times N}$. In this study the aim is to decompose $\bm{V}$ into a basis matrix $\bm{H}\in\mathbb{R}_+^{M\times K}$ and a weight matrix $\bm{W}\in\mathbb{R}_+^{K\times N}$, where $M$ denotes the number of features, $N$ denotes the number of observations, and $K$ denotes the number of basis vectors in the latent representation of $\bm{V}$. A schematic overview of all considered decompositions in this section can be seen in Figure \ref{fig:AE_sceheme}.
\subsection{Non-negative matrix factorization}
NMF decomposes a matrix with non-negative entries into a matrix product of two factor matrices with non-negative entries, one containing a set of basis vectors and one containing a set of weights. The shared dimension, $K$, of the factor matrices, is typically chosen to be much smaller than the dimensions of the input matrix, making NMF a dimensionality reduction technique.

Standard $K$-dimensional NMF was introduced by \citet{lee1999} and aims to make a reconstruction, $\bm{\hat{V}}$, of the original data matrix by a product of two non-negative matrices:
\begin{equation}\label{eq:NMF}
    \bm{\hat{V}} = \bm{HW},
\end{equation}
where each column in $\bm{H}$ represents a basis vector and each column in $\bm{W}$ represents each sample's weights when being reconstructed as a linear mixture of the basis vectors, i.e., \begin{equation}\label{eq:linrec}
    \bm{\hat{v}}_n = \sum_{k=1}^K \bm{h}_k w_{k,n},\quad n=1,\dots,N,
\end{equation} where $\bm{\hat{v}}_n$ is the $n$'th column of the reconstructed data matrix, $\hat{\bm{V}}$, $w_{n,k}$ represents the $(k,n)$'th entry of the weight matrix $\bm{W}$, and $\bm{h}_k$ represents the $k$'th column of the basis matrix $\bm{H}$ \citesq{lee1999}.

\subsection{Convex NMF}
Convex NMF, as introduced by \citet{cvxnmf}, is a special case of NMF, where the basis vectors are constrained to be spanned by the columns of the data matrix $\bm{V}$, thus the data matrix is approximated by:\begin{equation}\label{eq:cvxnmf}
    \bm{\hat{V}} = \bm{V}\bm{W}_1\bm{W}_2,
\end{equation}
with $\bm{W}_1,\bm{W}_2^T\in\mathbb{R}_+^{N\times K}$. Defining $\bm{H}:=\bm{VW}_1$ and $\bm{W}:=\bm{W}_2$ gives the model formulation of NMF as defined in Equation \eqref{eq:NMF}. \citet{cvxnmf} focus mainly on the general case where the data matrix, $\bm{V}$, can assume values within all real numbers, but for this paper we consider a version of convex NMF where $\bm{V}$ is constrained to be non-negative.

\subsection{Autoencoders}\label{sec:AE}
Autoencoders consist of an encoder applied to the input to create a latent representation and a decoder that maps the latent representation to a reconstruction of the input \citesq{Kramer1991NonlinearPC}. Choosing the dimension of the latent representation to be lower than the dimension of the input makes the autoencoder a dimensionality reduction technique. 
A single hidden layer and fully connected autoencoder's reconstruction, $\hat{\bm{V}}$, of a data matrix, $\bm{V}$, is mathematically defined as: \begin{equation}
\label{eq:AEdef}
   \hat{\bm{V}} = \phi_{\text{dec}}(\phi_{\text{enc}}(\bm{V}\bm{W}_{\text{enc}} + \bm{b}_{\text{enc}})\bm{W}_{\text{dec}} + \bm{b}_{\text{dec}}),
\end{equation}
where $\bm{W}_{\text{enc}}, \bm{W}_{\text{dec}}^T\in\mathbb{R}^{N\times K}$ are the encoding and decoding matrices, $\bm{b}_{\text{enc}}\in\mathbb{R}^{K}$ and $\bm{b}_{\text{dec}}\in\mathbb{R}^{N}$ are the bias terms of the encoding and the decoding layers and $\phi_{\text{enc}}:\mathbb{R}^{M\times K} \mapsto \mathbb{R}^{M\times K} $ and $\phi_{\text{dec}}:\mathbb{R}^{M\times N} \mapsto \mathbb{R}^{M\times N}$ are the activation functions which provide entry-wise modifications of the affected nodes.

\subsection{Mathematical Equivalence and Interpretation}
\label{sec:comp}
Setting $b_{\text{enc}} = \bm{0}_K$, $b_{\text{dec}} = \bm{0}_N$, and $\phi_{\text{enc}}, \phi_{\text{dec}}: x\mapsto x$ while constraining the weights to be non-negative in Equation \eqref{eq:AEdef} yields exactly the convex NMF formulation from Equation \eqref{eq:cvxnmf}, where $\bm{W}_1$ corresponds to the encoding matrix and $\bm{W}_2$ corresponds to the decoding matrix. Thus, non-negative autoencoders can be constructed to be mathematically equivalent to convex NMF. The choice between convex NMF and the autoencoder defined above therefore reduces to a choice of how to optimize the problem, either by the multiplicative updating steps derived by \citet{cvxnmf} or by the gradient descent-based additive updates of the autoencoder. The core architecture of this autoencoder is analogous to that of \citet{pei}, but differ by fixing $b_{\text{enc}}$ and $b_{\text{dec}}$ to zero instead of estimating them through training and choosing the identity function as activation functions instead of $\phi_{\text{enc}}:A\mapsto ReLU(x) = \text{max}(0,A)$ and $\phi_{\text{dec}}:A\mapsto Softmax(A) = \{\exp(a_{m,n})/\sum_{m = 1}^M \exp(a_{m,n})\}_{m = 1,\dots,M,n = 1,\dots,N}$, for a matrix $\bm{A}\in\mathbb{R}^{M\times N}$.

We will use 'C-NMF' to refer to convex NMF optimized with the multiplicative updating steps derived by \citet{cvxnmf} and use 'AE-NMF' for the class of autoencoders constructed equivalently to convex NMF.

The encoded data matrix in AE-NMF and C-NMF, $\bm{VW}_1$, which is a convex combination of the columns of the data matrix, is interpreted as the basis matrix, $\bm{H}$, in conventional NMF and hereby the interpretation of $\bm{W}_2$ corresponds to that of $\bm{W}$ in conventional NMF. Specifically, within mutational signature analysis this means that the signature matrix is a convex combination of the patients' mutational profiles which aligns with how mutational signatures are understood. If the data matrix is transposed to follow conventional orientation in neural networks, i.e., with patients as rows and mutation types as columns the weights or exposures are modelled as convex combinations of the mutations which lacks an equally straightforward interpretation and direct equivalence with convex NMF. The equivalence between C-NMF and AE-NMF is the necessary and previously missing link that enables one to interpret the parameters in AE-NMF similarly to NMF, therefore this orientation is crucial for proper comparison. 

Though the interpretation of AE-NMF, C-NMF, and NMF is similar, there are still considerable differences between C-NMF, AE-NMF, and standard NMF. In particular, NMF estimates $N\cdot K + K\cdot M$ parameters whereas AE-NMF and C-NMF estimates $2\cdot(K\cdot N)$ parameters. Thus, AE-NMF and C-NMF will estimate a larger number of parameters in the factor matrices than NMF when the number of observations $N$ surpasses the number of features $M$, which is often the case within mutational signatures.

\subsection{Modeling and performance framework}
\label{sec:est}
All estimation in this paper, for NMF, C-NMF, and AE-NMF, is done by minimizing the average Frobenius distance between the original input and the reconstructed input:
\begin{align}\label{eq:Frob}
    L_F(\bm{V},\bm{\hat{V}}) =  \frac{||\bm{V}- \bm{\hat{V}}||_F }{M\cdot N}= \frac{1}{M\cdot N}\sqrt{\sum_{i = 1}^M\sum_{j = 1}^N |v_{i,j}-\hat{v}_{i,j}|^2} .
\end{align} 
NMF is estimated using the multiplicative and iterative updating steps defined by \citet{lee2001}:
\begin{align}\label{eq:MUNMF}
\bm{H} \leftarrow \bm{H}\frac{\bm{VW}^T}{\bm{HWW}^T},\quad\quad \bm{W}\leftarrow\bm{W}\frac{\bm{H}^T\bm{V}}{\bm{H}^T\bm{H}\bm{W}}.
\end{align}
C-NMF is estimated using the multiplicative updating scheme defined by \citet{cvxnmf}. This updating scheme reduces to 
\begin{align}
\begin{split}\label{eq:update_cvx}
        \bm{W}_1 \leftarrow \bm{W}_1\sqrt{\frac{(\bm{V}^T\bm{V})\bm{W}_2^T}{(\bm{V}^T\bm{V})\bm{W}_1\bm{W}_2\bm{W}_2^T}}, \quad\quad
        \bm{W}_2^T  \leftarrow \bm{W}_2^T\sqrt{\frac{(\bm{V}^T\bm{V})\bm{W}_1}{\bm{W}_2^T\bm{W}_1^T(\bm{V}^T\bm{V})\bm{W}_1}},
\end{split}
\end{align}
in the case where $\bm{V}$ is constrained to be in the non-negative domain.

AE-NMF is estimated by an Adam optimizer, where the $t$'th iteration of a parameter, $w^{(t)}$, is updated additively and iteratively using the following method:\begin{align}\begin{split}
    w^{(t+1)}\leftarrow w^{(t)}-\eta \frac{\frac{m^{(t)}}{1-\beta_1^t}}{\sqrt{\frac{v^{(t)}}{1-\beta_2^t}} + \varepsilon}, \quad \text{where}\quad & m^{(t)} \leftarrow\beta_1m^{(t-1)} + (1-\beta_1)\nabla_w L_t \\ \text{and} \quad & v^{(t)}\leftarrow \beta_2 v^{(t-1)} + (1-\beta_2)(\nabla_w L_t)^2.
\end{split}
\end{align}
Here $L_t$ is the loss function at the $t$'th iteration and $\varepsilon= 10^{-8}$ is a small scalar to prevent division by zero. The factors $\beta_1$ and $\beta_2$ are exponential decay rates initialized to the default values $\beta_1 = 0.9$ and $\beta_2 = 0.999$. The moment vectors of parameter $w$ at time $t$, $m^{(t)}$ and $v^{(t)}$, are initialized as $m^{(0)} \leftarrow 0$ and $v{(0)}\leftarrow 0$ \citesq{adam}. Training of AE-NMF is done using the PyTorch module, version 1.13.1 \citesq{pytorch}, in Python version 3.11.3 \citesq{python3.11}. 
\subsubsection{Average Cosine Similarity}
The similarity between two sets of basis vectors is evaluated using the average cosine similarity (ACS). The cosine similarity for two equal length vectors, $\Tilde{\bm{h}},\hat{\bm{h}}\in\mathbb{R}^M$, is defined as\begin{equation}
    \label{eq:cs}
    S_C(\Tilde{\bm{h}},\hat{\bm{h}}) = \frac{\Tilde{\bm{h}}\cdot\hat{\bm{h}}}{||\Tilde{\bm{h}}||||\bm{\hat{h}}||}.
\end{equation}
The ACS for two matrices of matched basis vectors $\Tilde{\bm{H}},\hat{\bm{H}}\in\mathbb{R}_+^{M\times K}$ with $K$ basis vectors is defined as \begin{equation}\label{eq:ACS}
    ACS(\Tilde{\bm{H}}, \bm{\hat{H}}) = \frac{1}{K}\sum_{k = 1}^KS_c(\bm{\Tilde{h}}_k,\bm{\hat{h}}_k),
\end{equation} where $\Tilde{\bm{h}}_k$ denotes the $k$'th column in $\Tilde{\bm{H}}$, and  $\hat{\bm{h}}_k$ is the corresponding basis vector in $\hat{\bm{H}}$. 
\subsubsection{Signature Matching}
Given two matrices of basis vectors $\bm{H}\in\mathbb{R}^{M\times K}$ and $\Tilde{\bm{H}}\in\mathbb{R}^{M\times \tilde{K}}$ where $K\leq\Tilde{K}$, the task is to find $K$ pairs $(i,j)$, where $i = 1,\dots,K$ and $j \in\{1,\dots, \Tilde{K}\}$,  such that the sum of cosine similarities (and thus the ACS) over all pairs $\sum_{(i,j)}S_c(\bm{h}_i,\bm{\Tilde{h}}_j)$ is maximized, all vectors in $\bm{H}$ are matched exactly once, and all vectors in $\Tilde{\bm{H}}$ are matched at most once. This combinatorial problem is a linear assignment problem, where the cost matrix is all cosine distances between two basis vectors $\left\{1-S_c(\bm{h}_i,\Tilde{\bm{h}}_j)\right\}_{i = 1,\dots,K,\newline j = 1,\dots, \Tilde{K}}$. This problem is solved using the Hungarian algorithm in both the balanced ($K=\Tilde{K}$) and unbalanced ($K<\Tilde{K}$) case \citesq{Hungarian_alg}.

\section{Results}\label{sec:res}
In this section, we compare NMF, C-NMF, and AE-NMF in a simple simulated example (Section \ref{sec:sim}) and compare NMF and AE-NMF on the ovary, prostate, and uterus genomes from the Genomics England 100,000 Genomes Project \citesq{GEL,GEL2} by the ability to reconstruct the input data accurately and to generate stable and sensible mutational signatures (Section \ref{sec:cancer}).

All training was performed with a relative tolerance of $10^{-10}$ as convergence criteria. Estimation with AE-NMF was performed with an Adam optimizer with a learning rate of $10^{-4}$. Non-negativity in AE-NMF was enforced taking the absolute value of the encoding and decoding weight matrices in the forward pass. The choice of method to enforce non-negativity in AE-NMF is elaborated in Supplementary Material Section 1.4.
\subsection{Simulated example}
\label{sec:sim}
Consider an example with two basis vectors (signatures) consisting of 6 features (mutation types):
\begin{align}
\begin{split}
    \bm{h}_1 &= (2,2,1,1,0,0)^T/6\\
    \bm{h}_2 &= (0,0,0,1,1,1)^T/3.
\end{split}
\end{align}
From these basis vectors we simulated 30 samples (patients), $\{\bm{v}_1,\dots,\bm{v}_{30}\}$, where each sample was simulated with one of three distinct compositions of $\bm{h}_1$ and $\bm{h}_2$ and Poisson distributed noise:
\begin{align}
    \bm{v}_j = (v_{j,1},v_{j,2},v_{j,3},v_{j,4},v_{j,5},v_{j,6})^T,\end{align}
for $j=1,\dots, 30$ and 
\begin{align}
    v_{i,j} \overset{\text{i.i.d}}{\sim} \begin{cases}
          \text{Po}(180\cdot \bm{h}_{i,1} + 20\cdot \bm{h}_{i,2}) &\text{for } j = 1,\dots, 10\\ 
          \text{Po}(100\cdot \bm{h}_{i,1} + 100\cdot \bm{h}_{i,2}) &\text{for } j = 11,\dots, 20\\
          \text{Po}(20\cdot \bm{h}_{i,1} + 180\cdot \bm{h}_{i,2}) &\text{for } j = 21,\dots, 30,\\
     \end{cases}
\end{align}
for $i = 1,\dots,6$. For the full data matrix, $\bm{V} = \left[v_{i,j}\right]$, the two basis vectors and the corresponding weights (exposures) were estimated using NMF, C-NMF, and AE-NMF, with the Frobenius norm as the loss function. To encourage convergence towards the same minimum, C-NMF and AE-NMF were initialized with the same matrices, $\bm{W}_1^{(0)}$ and $\bm{W}_2^{(0)}$, respectively to a $30\times 2$ and  a $2\times 30$ matrix with values sampled uniformly in $[0,1)$. This was not possible for NMF since its factor matrices have different dimensions. For NMF the matrix of basis vectors was initialized uniformly in $[0,1)$, and the weight matrix was initialized as $\bm{W}_2^{(0)}$ from C-NMF and AE-NMF.
\begin{figure}
    \centering
    \includegraphics[width =0.75 \linewidth]{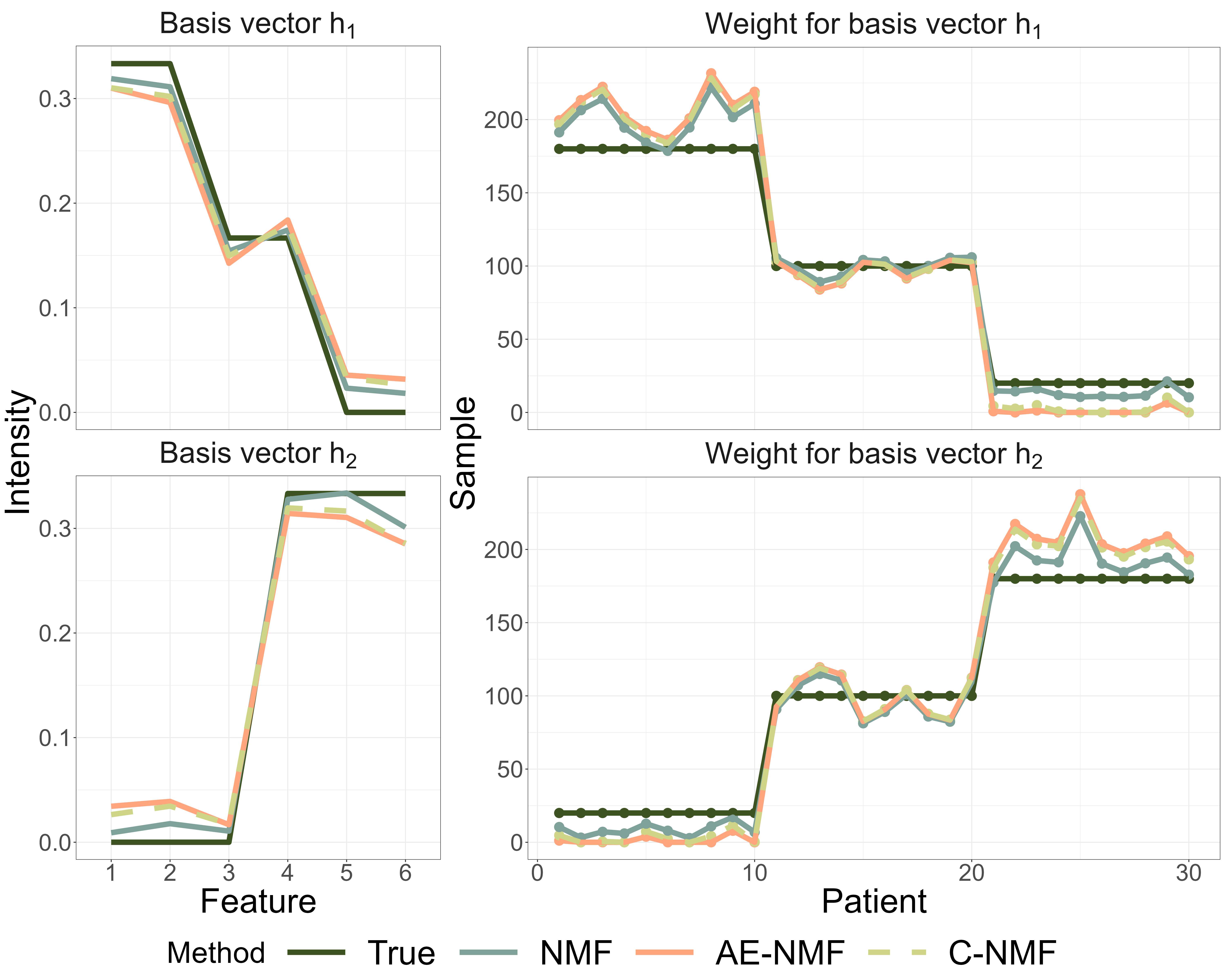}
    \caption{Estimated and true weights and basis vectors of the simulated data. Note the light green lines representing C-NMF coincide with the the dashed, coral lines of AE-NMF since the estimates found by both methods are almost identical.}
    \label{fig:sim_est}
\end{figure}

Figure \ref{fig:sim_est} shows the estimated basis vectors and weights for each method. In this example, identical factor matrices are recovered by the C-NMF and AE-NMF updates, illustrated by the coinciding dashed green and coral lines, while NMF yields different results, illustrated by the blue lines. Furthermore, the NMF solution reconstructs the input with higher accuracy, with a reconstruction error of $12.95$ compared to the higher values of $13.64$ for AE-NMF and $13.00$ for C-NMF.

\subsection{Cancer Data Analysis}
\label{sec:cancer}
To emulate the terminology used in the cancer data analysis, the matrix of basis vectors will be denoted as the signatures, and the weight matrix will be denoted the exposures.

The number of signatures in each diagnosis was determined with the method described in the Supplementary Material Section 1.3 with ten training/test splits for each diagnosis. The test errors as a function of $K\in \{2,\dots,12\}$ are depicted in Figure S1. Algorithm S2 in the Supplementary Material yielded three signatures for AE-NMF and four signatures for C-NMF and NMF in the ovary cohort; four, four and six signatures for AE-NMF, C-NMF and NMF, respectively in the prostate cohort and four, six, and 11 signatures for AE-NMF, C-NMF, and NMF respectively in the uterus cohort. Thus, the number of signatures used in the cancer data analyses is chosen, using the weighted average in Equation (S4) in the Supplementary Material, as four for the ovary cohort, five for the prostate cohort and eight for the uterus cohort.

As shown in Section \ref{sec:comp} C-NMF and AE-NMF are mathematically equivalent. Furthermore, since the resulting factor matrices of C-NMF and AE-NMF were identical in the simulated example and performed similarly in the cancer data analysis with respect to reconstruction error (Figure S1 and Figure S3) and consistency (Figure S4), we consider C-NMF and AE-NMF as practically equivalent. Thus, the following analyses will be performed comparing only NMF and AE-NMF.

\subsubsection{Extraction performance}
\label{sec:acc}
\begin{figure}
  \centering\includegraphics[width=\linewidth]{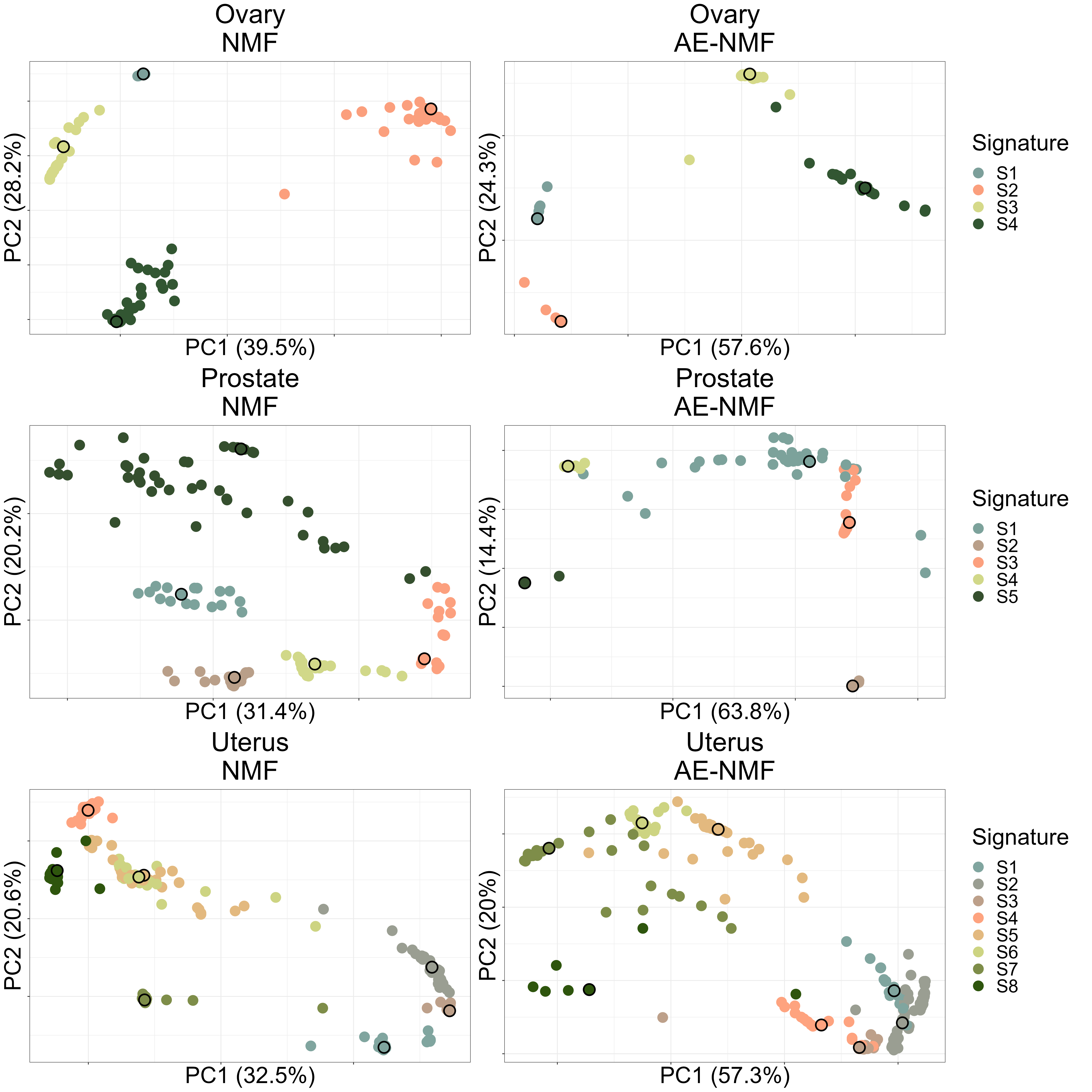}\par
  \caption{Second principal component plotted agains the first principal component of the \textit{de novo} extracted signatures from the 30 train/test splits for AE-NMF and NMF (columns) and each diagnosis (rows). Points are colored by the PAM clustering assignment, and the cluster mediod is highlighted with a black outline. NMF: non-negative matrix factorization; AE: autoencoder; PAM: partition around mediods.}
  \label{fig:PC6}
\end{figure}

For each cohort, we divided the patients' mutational profile data into 30 80/20 train/test set splits. \textit{De novo} extractions by NMF  and AE-NMF were performed on the training set yielding a set of training errors and refits were performed on the test sets yielding a set of test errors. Plots of the first and second principal component of all extracted signatures from the 30 training sets are depicted in Figure \ref{fig:PC6}. 

\begin{figure}
    \centering
    \includegraphics[width = \linewidth]{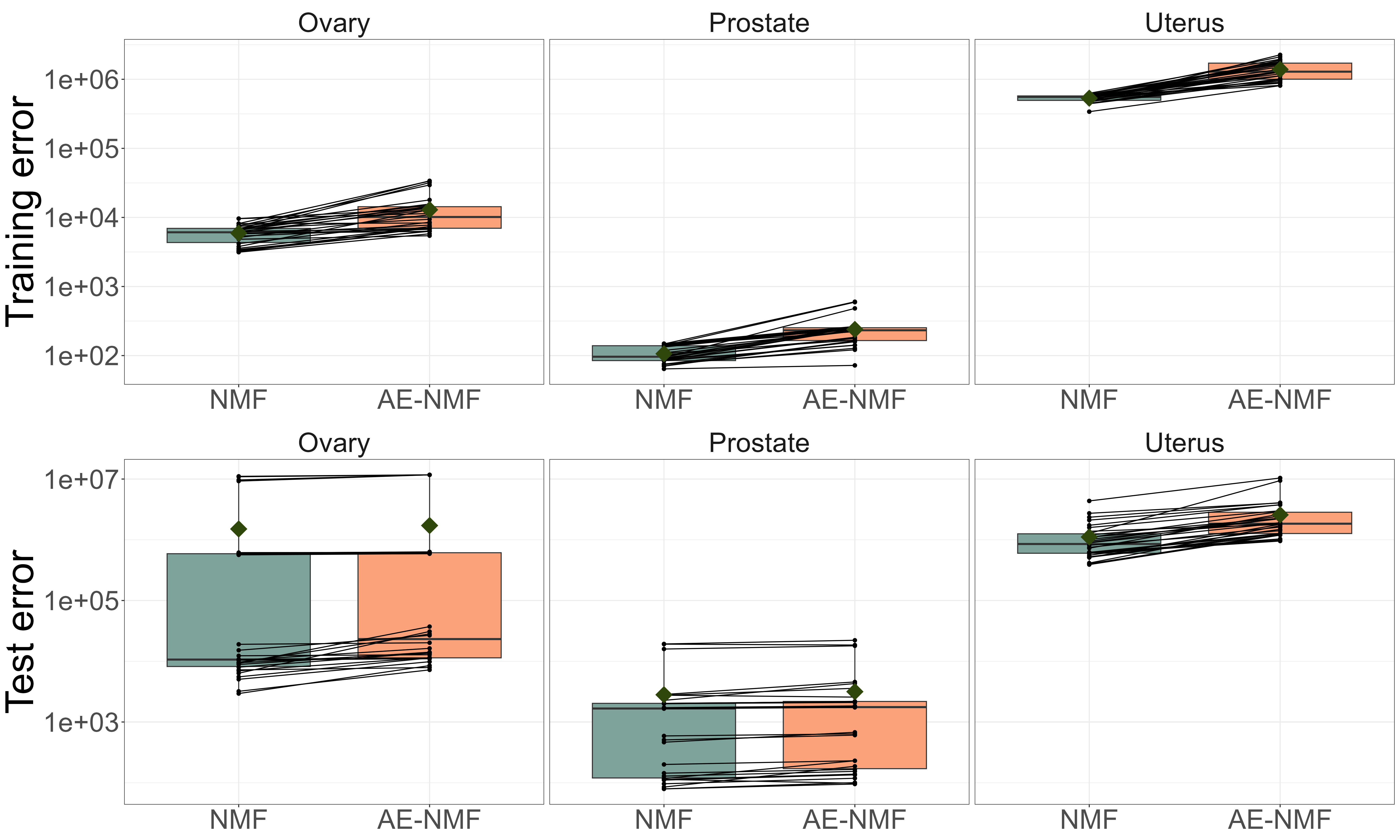}
    \caption{Boxplots of the training and test errors of 30 train/test splits of the ovary, prostate and uterus cohorts. NMF and AE-NMF errors resulting from the same splits are connected by a black line. The boxes are colored corresponding to the method used, and a green diamond depicts the average error. The y-axis is on log 10 scale.}
    \label{fig:error}
\end{figure}

The procedure of calculating the test errors is detailed in the Supplementary Material Section 1.2. Boxplots of the training and test errors when reconstructing the input matrix for each method and diagnosis are shown in Figure \ref{fig:error}. Table \ref{tab:perf} reports the average training and test error across the 30 splits for each method and diagnosis.
Considering the reconstruction errors in Figure \ref{fig:error} and Table \ref{tab:perf}, it is apparent that NMF consistently performs better than AE-NMF in terms of reconstructing the input data. This is the case both on average and in the majority of splits as shown by Figure \ref{fig:error}. The ratios in Table \ref{tab:perf} reveal that the difference is more expressed in the training splits than the test splits.

\begin{figure}
    \centering
    \includegraphics[width = \linewidth]{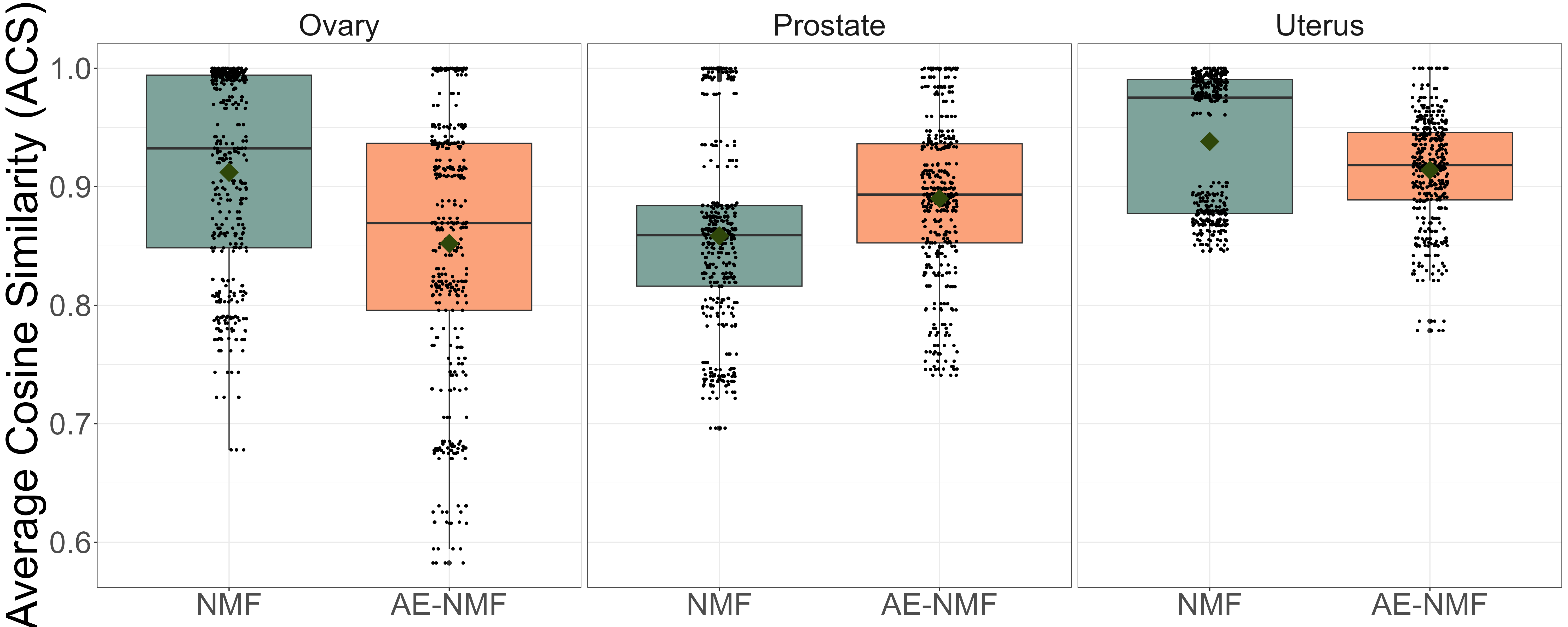}
    \caption{Boxplots of the ACS between each combination of signatures, extracted using NMF or AE-NMF and across the 30 splits of the data matrix for the ovary, prostate, and uterus cohort. The average is marked by a green diamond.}
    \label{fig:cons}
\end{figure}

\begin{table}[]
\centering
\begin{tabular}{llrrrr}
                                                   &                      &                    & \multicolumn{2}{c}{\textbf{Average Error}}    &            \\
\textbf{Cohort}                                    & \textbf{Split}       & $\bm{n}$           & \textbf{NMF}             & \textbf{AE-NMF}    & \textbf{Ratio}     \\ \hline
\multirow{2}{*}{\textbf{Ovary}}                    & Train                & $418$              & $\bm{5.92\cdot 10^{3}}$   & $1.29\cdot 10^4$   &  $2.17$ \\
                                                   & Test                 & $105$              & $\bm{1.51\cdot 10^{6}}$   & $1.72\cdot 10^6$   &  $1.63$\\ \hline
\multirow{2}{*}{\textbf{Prostate}}                 & Train                & $248$              & $\bm{1.06\cdot 10^2}$     & $2.39\cdot10^2$    & $2.24$\\
                                                   & Test                 & $63$               & $\bm{2.81\cdot 10^{3}}$   & $3.16\cdot 10^3$ &  $1.23$\\ \hline
\multirow{2}{*}{\textbf{Uterus}}                   & Train                & $570$              & $\bm{5.33\cdot 10^{5}}$   & $1.39\cdot 10^6$ & $2.61$  \\
                                                   & Test                 & $143$              & $\bm{1.11\cdot 10^{6}}$   & $2.57\cdot 10^{6}$ &	$2.45$  \\

\end{tabular}
\caption{Average training and test error between the input and reconstructed data for each method across the 30 splits of the ovary, prostate, and uterus cohort and the average ratio between the errors. The lowest reconstruction error in each cohort is highlighted in bold.}
\label{tab:perf}
\end{table}
For each method and diagnosis the analyses yielded 30 signature sets, one from each training set. The consistency of the estimated signatures extracted within each method is investigated by calculating the ACS between each pairwise combination of signatures extracted using a given method across the 30 splits of the data matrix. This yields a total of $\binom{30}{2} = 435$ comparisons for each cohort and method, and resulted in an average ACS consistency of $0.91$ for NMF and $0.85$ for AE-NMF in the ovary cohort, $0.86$ for NMF and $0.88$ for AE-NMF in the prostate cohort, and $0.94$ for NMF and $0.91$ for AE-NMF in the uterus cohort. To asses whether the difference in mean consistency between NMF and AE-NMF were significant, two-sided t-tests for equal means were performed for each cohort. These resulted in a $p$-value of $3.7\cdot10^{-18}$ for the ovary cohort, $7.0\cdot 10^{-10}$ for the prostate cohort, and $1.1^\cdot 10^{-11}$ for the uterus cohort, thus the mean consistency for NMF and AE-NMF can not be assumed to be equal for any cohort. The consistencies are depicted by boxplots in Figure \ref{fig:cons}, with green diamonds marking the average consistencies listed above. This reveals generally comparable consistencies of NMF and AE-NMF signatures.
\subsubsection{COSMIC Validation}
To compare the estimated signatures to those of the leading library of mutational signatures, COSMIC v. 3.4 \citesq{cosmic_2020}, the signatures were clustered to form sets of consensus signatures using the partitioning around medoids (PAM) algorithm \citesq{PAM} with the number of clusters equal to the number of signatures used in the initial extractions. The clustering is depicted in Figure \ref{fig:PC6} where the first and second principal components of all \textit{de novo} signatures are depicted and points are colored by their assigned PAM clustering. The consensus signatures were subsequently matched to the COSMIC signatures. The matched COSMIC signatures along with their cosine similarity can be seen in Table \ref{tab:COSMIC}.\newline
\begin{table}[]
\centering
\begin{tabular}{llllll}
\textbf{Cohort} & \textbf{Signature} & \multicolumn{2}{c}{\textbf{NMF}} & \multicolumn{2}{c}{\textbf{AE-NMF}} \\ \hline
\multirow{5}{*}{\rotatebox{90}{\textbf{Ovary}}}    & S1  & SBS10a           & 0.93          & SBS10a            & 0.93            \\
                                   & S2  & \textbf{SBS3}           & 0.95          & SBS40a            & 0.80            \\
                                   & S3  & SBS10c           & 0.70          & SBS10c            & 0.74            \\
                                   & S4  & SBS44            & 0.84          & SBS44             & 0.82            \\ \cline{2-6}
                                   & ACS &                  & 0.80          &                   & 0.82            \\ \hline 
\multirow{6}{*}{\rotatebox{90}{\textbf{Prostate}}} & S1  & SBS17b           & 0.84          & SBS17b            & 0.77            \\
                                   & S2  & \textbf{SBS33}            & 0.95          & \textbf{SBS33}             & 0.93            \\
                                   & S3  & \textbf{SBS1}             & 0.94          & \textbf{SBS6}              & 0.90            \\
                                   & S4  & SBS44            & 0.86          & SBS44             & 0.85            \\
                                   & S5  & SBS40a           & 0.90          & SBS40a            & 0.87            \\\cline{2-6}
                                   & ACS &                  & 0.90          &                   & 0.86            \\ \hline 
\multirow{9}{*}{\rotatebox{90}{\textbf{Uterus}}}   & S1  & SBS10d           & 0.96          & SBS10d            & 0.95            \\
                                   & S2  & SBS10c           & 0.84          & \textbf{SBS10a}            & 0.95            \\
                                   & S3  & \textbf{SBS10a}          & 0.97          & SBS10c            & 0.75            \\
                                   & S4  & \textbf{SBS6}             & 0.88          & \textbf{SBS6}              & 0.84            \\
                                   & S5  & \textbf{SBS20}            & 0.93          & \textbf{SBS20}             & 0.90            \\
                                   & S6  & \textbf{SBS10b}           & 0.78          & \textbf{SBS2}              & 0.71            \\
                                   & S7  & \textbf{SBS28}            & 0.96          & \textbf{SBS28}            & 0.70            \\
                                   & S8  & \textbf{SBS26}            & 0.83          & \textbf{SBS26}             & 0.82            \\\cline{2-6}
                                   & ACS &                  & 0.89          &                   & 0.83           
\end{tabular}
\caption{The COSMIC signatures matched to the consensus signatures and the corresponding cosine similarity. The average cosine similarity (ACS) between a consensus signature set and the matched COSMIC signatures is reported in the last line for each diagnosis. Signatures that have previously been observed in patients with the same diagnosis are highlighted with bold text.}
\label{tab:COSMIC}
\end{table}
The ACS between the consensus signatures and their matched COSMIC signatures is 0.80 for NMF and 0.82 for AE-NMF in the ovary cohort, 0.90 for NMF and 0.86 for AE-NMF in the prostate cohort and 0.90 for NMF and 0.83 for AE-NMF in the uterus cohort. This reveals a slightly higher average similarity to COSMIC for NMF than AE-NMF. Additionally, the majority of NMF signatures also have better COSMIC matches than the corresponding AE-NMF signature. The matched COSMIC signatures for all splits before clustering for NMF and AE-NMF can be seen in Supplementary Figure S5-S7. Of all matched consensus signatures, the following proportion has been previously observed in the corresponding diagnosis for each cohort: 1/4 of NMF and 0/4 of AE-NMF signatures in the ovary cohort, 2/5 of both NMF and AE-NMF signatures in the prostate cohort, and 6/8 of both NMF and AE-NMF signatures in the uterus cohort. 

Overall both NMF and AE-NMF show a high degree of conformity with the COSMIC SBS signatures, and the extracted signatures are relevant to the diagnoses in which they have been identified. In signature cosistency, conformity with COSMIC and choosing relevant signatures, NMF and AE-NMF perform similarly, perhaps with a slight advantage to NMF.

\section{Discussion}\label{sec:disc}
In this study, we compare NMF and AE-NMF by their ability to extract valid and consistent basis vectors and creating accurate reconstructions of the input data. We assert that such comparisons are theoretically meaningful since we demonstrate that AE-NMF and C-NMF are mathematically equivalent.

The study focuses on extracting mutational signatures in the ovary, prostate, and uterus cancer genomes of Genomics England's 100,000 Genomes cohort. NMF consistently outperformed AE-NMF in terms of reconstruction error; the differences being more expressed in the training splits than in the test splits. While AE-NMF constrains parameters to convex combinations of patients' profiles, NMF can freely assume non-negative values, giving it an advantage in training set reconstruction. When reconstructing the test set the signature matrix is fixed, and the task is thus identical for NMF and AE-NMF. One could expect the constrained nature of AE-NMF to regularize the signatures such that they reconstruct the test splits better compared to NMF, but as this was not the case in this study, it suggests that AE-NMF signatures may be generally less informative than the corresponding NMF signatures. The COSMIC validation revealed that both models recovered relevant signatures with high cosine similarity, with a slight advantage to NMF. This advantage is likely driven by the fact that the majority of COSMIC signatures are extracted using NMF-based methods \citesq{alexandrov2020}. 

The mathematical equivalence between convex NMF and non-negative autoencoders enables the interpretation of parameters from AE-NMF to be similar to that of NMF. Thus, the mathematical equivalence between convex NMF and non-negative autoencoders identified in this study is a necessary link in properly comparing AE-NMF and NMF. A link that has been missing in previous attempts to replace NMF with autoencoders while using the same interpretation. \citet{squires} also came to the conclusion that an autoencoder with the architecture of AE-NMF yields a hidden layer consisting of convex combinations of the data points but did not make the connection to convex NMF.

In practice we observed that C-NMF and AE-NMF will yield identical solutions in a sufficiently simple setup. When increasing the complexity of the problem both methods perform similarly in terms of reconstruction error and stability of the basis vectors, albeit finding different solutions within this minimum.

The architecture of AE-NMF is atypical for autoencoders by its shallow and linear nature and by transposing the input data matrix. By orienting the input matrix, $\bm{V}$, with features ($M$) as rows and observations ($N$) as columns the architecture will stray from how data is conventionally passed through neural networks, but this orientation is not uncommon in the literature of non-negative autoencoders \citesq{Khatib, pei, squires}. One direct consequence of this orientation is that it will not be possible to fit a separate sample by passing it through the trained network, a compelling attribute of neural networks, as the trained parameters will be observation-specific. Furthermore, the shallowness of AE-NMF favours the capture of linear relationships over more complex patterns. It is these strict architectural choices that yields the equivalence to convex NMF, ensuring a precise understanding of the interpretation of the parameters. Other efforts, such as the MUSE-XAE autoencoder for mutational signature extraction \citesq{MUSE-XAE}, have utilized the benefits of classic autoencoders with deeper extraction and conventional orientation of the input data matrix but at the cost of jeopardizing the link to NMF and thus the exact interpretation of the parameters, since this autoencoder is not equivalent to C-NMF. 

All methods were optimized by minimizing the Frobenius loss function since this is the only loss function for conventional convex NMF by \citet{cvxnmf} that presently has updating steps. This loss is only asymptotically efficient given Gaussian distributed input data, which is a questionable assumption since the data considered in this study consists of  count data. Choosing a loss function adapted to Poisson may therefore be more adequate for this problem \citesq{alexandrov2020,GEL0}. An even more sophisticated error model would be the Negative Binomial distribution \citesq{SigMOS}. The Negative Binomial distribution can model overdispersion in the mutational counts, which is often present on the patient-specific level. In future work, it would be interesting to see how training with a more appropriate loss function affects how NMF and AE-NMF compare in reconstruction accuracy and stability and whether the results from this study generalize to the Kullback-Leibler divergence. 

In this study a relative convergence criteria of $10^{-10}$ was used for all analyses. Using such a low convergence criteria promoted that the three methods, based on two vastly different updating schemes, converged towards similar minima and, thus, established a common ground for comparison. A lower relative tolerance in the real data analyses was computationally infeasible. In particular, the high tolerance made especially C-NMF extremely time consuming in cases with many patients and/or signatures. The bootstrap method for determining the number of signatures were limited to ten splits for the same reason. If the aim is to just extract signatures using either method, we do not necessarily recommend training with such a low tolerance.

\section{Conclusion}\label{sec:conc}
This study compares NMF with non-negative autoencoders, facilitated by the mathematical equivalence between non-negative autoencoders and convex NMF. This bridges a crucial gap in the comparison of NMF and non-negative autoencoders by offering insights into parameter interpretation that were previously lacking. 
The choice between convex NMF and its autoencoder equivalent is a question of choosing between a multiplicative or gradient descent-based optimizing algorithm to solve the same optimization problem. Thus, the non-negative autoencoder described in this study can be used as a faster alternative to convex NMF. Non-negative autoencoders exhibit higher reconstruction errors and similar consistencies compared to NMF, therefore the non-negative autoencoder investigated in this study is not a suitable alternative to NMF in mutational signature extraction. Thus, this study underscores the significance of methodological considerations when replacing NMF with non-negative autoencoders. On the other hand autoencoders hold promise for modeling non-linearity, a capability absent in NMF, but such advancements are made at the cost of exact parameter interpretation.

\section*{Acknowledgements}
\noindent This work was supported by the Danish Data Science Academy, which is funded by the Novo Nordisk Foundation (NNF21SA0069429) and VILLUM FONDEN (40516), the Novo Nordisk Foundation (NNF21OC0069105), and the Research Hive "REPAIR", which is funded by Aalborg University Hospital.

\section*{Code- and Data avaliability}
Code used for this study can be found at \href{https://github.com/CLINDA-AAU/AE-NMF}{Github} and the Genomics England WGS data used in this study was provided by \citet{GEL2} on \href{https://zenodo.org/records/5571551}{Zendo}.

\section*{Supplementary Materials}

\subsection*{Supplementary Methods}

The supplementary methods contains a section on how test errors are calculated for all methods, the algorithms for choosing the optimal number of basis vectors, and the theoretical overview of methods for enforcing non-negativity in the autoencoder.

\subsection*{Supplementary Results}

The supplementary results contains the results used to choose the optimal number of basis vectors, the results used to determine the method to constrain non-negativity, and additional figures and tables.


\printbibliography
\end{document}


\author[1,2]{Ida Egendal}
\author[1,2]{Rasmus Froberg Brøndum}
\author[3]{Marta Pelizzola}
\author[3]{Asger Hobolth}
\author[1,2]{Martin Bøgsted}

\affil[1]{Center for Clinical Data Science, Aalborg University and Aalborg University Hospital, Aalborg, Denmark}
\affil[2]{Clinical Cancer Research Center, Aalborg University Hospital, Aalborg, Denmark}
\affil[3]{Department of Mathematics, Aarhus University, Aarhus, Denmark}

\date{}

\title{Supplementary Material for On the Relation Between Autoencoders and Non-negative Matrix Factorization, and Their Application for Mutational Signature Extraction}
\maketitle

\section{Supplementary Methods}
\subsection{Exposure Frobenius Distance}
The difference between two matched weight matrices $\bm{\Tilde{W}},\bm{\hat{W}}\in\mathbb{R}_+^{K\times M}$ is measured as the average Frobenius distance between the two matrices:\begin{align}\label{eq:ED}
    L_F(\bm{\Tilde{W}},\bm{\Hat{W}}) = \frac{1}{K\cdot M}||\bm{\Tilde{W}} - \bm{\Hat{W}}||_F,
\end{align}
where the weights are rearranged to match using the signature matching found in the corresponding signature sets.
\subsection{Test error}
\label{sec:refit}
The test error is calculated as the loss between the test data and its reconstruction. First, we divide the data into a training dataset $\bm{V}_{\text{train}}$ and a test dataset $\bm{V}_{\text{test}}$. The basis matrix $\bm{\hat{H}}$, and the weight matrix, $\bm{\hat{W}}_{\text{train}}$, of the training dataset are estimated. Then, for the test dataset the basis matrix from the training set, $\bm{\hat{H}}$, is fixed and the weight matrix is estimated using non-negative least squares \citesq{NNLS}:
\begin{equation}
    \bm{\hat{W}}_{\text{test}} =\arg\min_{\bm{W}\geq 0}||\bm{\hat{H}}\bm{W} - \bm{V}_{\text{test}}||_F,
\end{equation} 
from which the reconstructed test dataset is defined as $\bm{\hat{V}}_{\text{test}}:=\bm{\hat{H}}\bm{\hat{W}}_{\text{test}}$. This representation is the test dataset reconstructed with the basis found in the training data. Now, the test error is calculated as
\begin{equation}\label{eq:in-out}
L_F(\bm{V}_{\text{test}},\bm{\hat{V}}_{\text{test}})= \frac{1}{M\cdot N}||\bm{V}_{\text{test}}- \bm{\hat{V}}_{\text{test}}||_F.
\end{equation}
\subsection{Choosing the number of basis vectors}
  \label{sec:chooseK}
\label{sec:bootstrap}
Let $K$ denote the number of basis vectors used for extraction.
The test error for each method and $K \in\{ 2,\dots, \mathcal{K}\}$ is calculated using Algorithm S\ref{alg:bootstrap K} which is similar to the SigMos algorithm \citesq{SigMOS}. The main difference being that Algorithm S\ref{alg:bootstrap K} samples with replacement and SigMos samples without replacement. \newline
 We consider the appropriate value of $K$ to be the value of $K$, where the test error does not change significantly when increasing $K$. To obtain this, iterative tests are performed on whether the test errors for a given $K$ can be assumed to have the same distribution as the test errors from the following $K+1$. This iterative procedure continues until a test for similar distributions does not fail anymore and $K$ at non-failure values is determined as the optimal number of basis vectors. \newline
Since the same train/test splits are used to calculate the test errors for each $K$, a two-sample paired Wilcoxon's test is used to test whether the sample from a given $K$ can be assumed to be distributed as the sample from $K+1$. A p-value of $p = 0.05$ is used as significance level. The exact procedure can be seen in Algorithm S\ref{alg:wilcox}.

To ensure comparability in the quantitative analyses, we choose to use the same number of basis vectors for all methods. In cases where Algorithm S\ref{alg:wilcox} chooses different $K$'s for each method, the final number of basis vectors is determined the weighted average of each method rounded to the nearest integer:
\begin{equation}
\label{eq:wavg}
    K_{\text{all}}  =\left[ \frac{K_{\text{NMF}} + \frac{1}{2}K_{\text{C-NMF}} + \frac{1}{2}K_{\text{AE-NMF}}}{2}\right],
\end{equation} with AE-NMF and C-NMF weighted by 1/2, since these are more likely to agree on the number of bases as they are variants of the same base model.
\begin{algorithm}[]
\label{alg:bootstrap K}
\SetAlgoLined
 Input $\bm{V}\in\mathbb{R}_+^{M\times N}$ and $nsims$\;

 \For{$i\in\{1,\dots, nsims\}$}{
 Sample $\text{idx}_{\text{train}}$ from $\{1,\dots,N\}$ with replacement\;
 $\bm{V}_{\text{train}} = \bm{V}[,\text{idx}_{\text{train}}]\in\mathbb{R}_+^{M\times N_{\text{train}}}$\;
 $\bm{V}_{\text{test}} = \bm{V}[,-\text{idx}_{\text{train}}]\in\mathbb{R}_+^{M\times N_{\text{test}}}$\;
 \For{model $\in \{NMF, C-NMF, AE-NMF\}$ and  $K \in \{2,\dots,\mathcal{K}\}$}{
    Fit $\bm{\Hat{H}},\bm{\Hat{W}}_{\text{train}}$ using $K$-dimensional model on $\bm{V}_{\text{train}}$\;
    Fit $\bm{\Hat{W}}_{\text{train}} = \arg\min_{\bm{W}\geq 0}||\bm{\hat{H}}\bm{W} - \bm{V}_{\text{test}}||^2_F$ \;
    Estimate $\bm{\Hat{V}_{\text{test}}} = \bm{\Hat{H}}\bm{\Hat{W}}_{\text{test}}$\;
    Calculate $x_{i, K, model} = L_F(\bm{V}_{\text{test}}, \bm{\Hat{V}}_{\text{test}})/(M\cdot N_{\text{test}})$\;
}}
\Return $\{x_{i, K, model}\}$ for $i\in\{1,\dots,nsims\}$, $K\in\{2,\dots,\mathcal{K}\}$ and $model\in \{NMF, C-NMF, AE-NMF\}$
 \caption{Test error as a function of number of basis vectors.}
 \end{algorithm}

\begin{algorithm}[]
	\label{alg:wilcox}
\SetAlgoLined

 Input test error sample $x_{1,K},\dots, x_{nsims,K}$ for $K = 2,\dots, \mathcal{K}$\;
 \For{$K\in\{2,\dots, \mathcal{K}-1\}$}{
     Calculate $p$ as the p-value of Wilcoxon's two sample paired test of $\{x_{i,K}\}_{i = 1,\dots,nsims}$ and  $\{x_{i,K+1}\}_{i = 1,\dots,nsims}$\;
     \If{$p<p_{val}$}{
     Return $K$
     }
     \Else{Continue}
 }
 Return $K$
 \caption{Iterative tests for the appropriate number of basis vectors.}
 \end{algorithm}

\subsection{Non-negativity in AE-NMF}
\label{seq:nonneg}

\begin{table}[]
\centering
\begin{tabular}{l|l|l}
Name           & Formula                                                                                                                                                                                                              & Source \\ \hline
ReLU  + PG     & $ReLU(\bm{VW}_{\text{enc}})|\bm{W}_{\text{dec}}|$                                                                                                                                                                    & \cite{PGNMF}  \\\hline
ReLU + SM      & $Softmax(ReLU(\bm{VW}_{\text{enc}}\bm{W}_{\text{dec}}))$                                                                                                                                                             & \cite{pei}    \\\hline
EGD            & $\bm{W}_{t+1} = \bm{W}_t\exp(-\eta \frac{ \partial L_t }{\partial \bm{W}_t})$                                                                                                                                        & \cite{Khatib} \\\hline
MU             & \makecell{$ \bm{W}_{\text{dec}, nk}^{(t+1)} = \bm{W}_{\text{dec}, nk}^{(t)}\frac{(\bm{V}(\bm{W}_{\text{enc}}\bm{V})^T)_{nk}}{(\bm{W}_{\text{dec}}^{(t)}\bm{W}_{\text{enc}}\bm{V}(\bm{W}_{\text{enc}}\bm{V})^T)_{nk}}$ \\
                            $\bm{W}_{\text{enc}, kn}^{(t+1)} = \bm{W}_{\text{enc}, kn}^{(t)}\frac{((\bm{W}_{\text{dec}}^T\bm{V})\bm{V}^T)_{kn}}{((\bm{W}_{\text{dec}}^T\bm{W}_{\text{dec}}\bm{W}_{\text{enc}}^{(t)}\bm{V})\bm{V}^T)_{kn}}$}         & \cite{AE_MU_thesis}  \\\hline
PG/FP PG& $\bm{V}ReLU(\bm{W}_{\text{enc}})ReLU(\bm{W}_{\text{dec}})$                                                                                                                                                           & -      \\\hline
Abs/FP Abs            & $\bm{V}|\bm{W}_{\text{enc}}||\bm{W}_{\text{dec}}|$                                                                                                                                                                   & -      \\\hline
\end{tabular}
\caption{Overview of non-negative constraining methods for autoencoders.}
\label{tab:nn}
\end{table}

There exist several ways of enforcing non-negativity in AE-NMF. An overview of these considered in this paper is provided in Table S\ref{tab:nn}. \citet{pei} suggested a ReLU activation function on the encoding layer and a softmax activation function on the decoding layer (ReLU + SM). \citet{AE_MU} found that using ReLU on the encoding layer while using a projected gradient on $\bm{W}_{\text{dec}}$ promoted convergence (ReLU + PG) \cite{PGNMF}. Other methods enforce non-negativity with multiplicative updating steps.  \citet{Khatib}, e.g., suggested updating using exponential gradient descent and \citet{AE_MU_thesis} suggested multiplicative updates using a cleverly constructed learning rate. These methods are all excluded in this study, since methods enforcing non-negativity using activation functions on the network layers instead of the weight matrices (ReLU + PG and ReLU + SM) allow $\bm{W}_{\text{enc}}$ and/or $\bm{W}_{\text{dec}}$ to assume negative values, losing the equivalence with convex NMF. Methods based on multiplicative updates (EGD and MU) are also excluded in order to focus on the conventional additive updating scheme of autoencoders to contrast the multiplicative updating scheme of C-NMF and NMF

This study considers a projected gradient on both $\bm{W}_{\text{enc}}$ and $\bm{W}_{\text{dec}}$ either after each update (PG) or as a part of the forward pass (FP PG), and taking the absolute values of $\bm{W}_{\text{enc}}$ and $\bm{W}_{\text{dec}}$ after each update (Abs) or as a part of the forward pass (FP Abs).

\section{Supplementary Results}
\label{FourthAppendix}
\subsection{Non-negativity in the autoencoder}
\begin{figure}
     \centering
     \includegraphics[width = 0.8\linewidth]{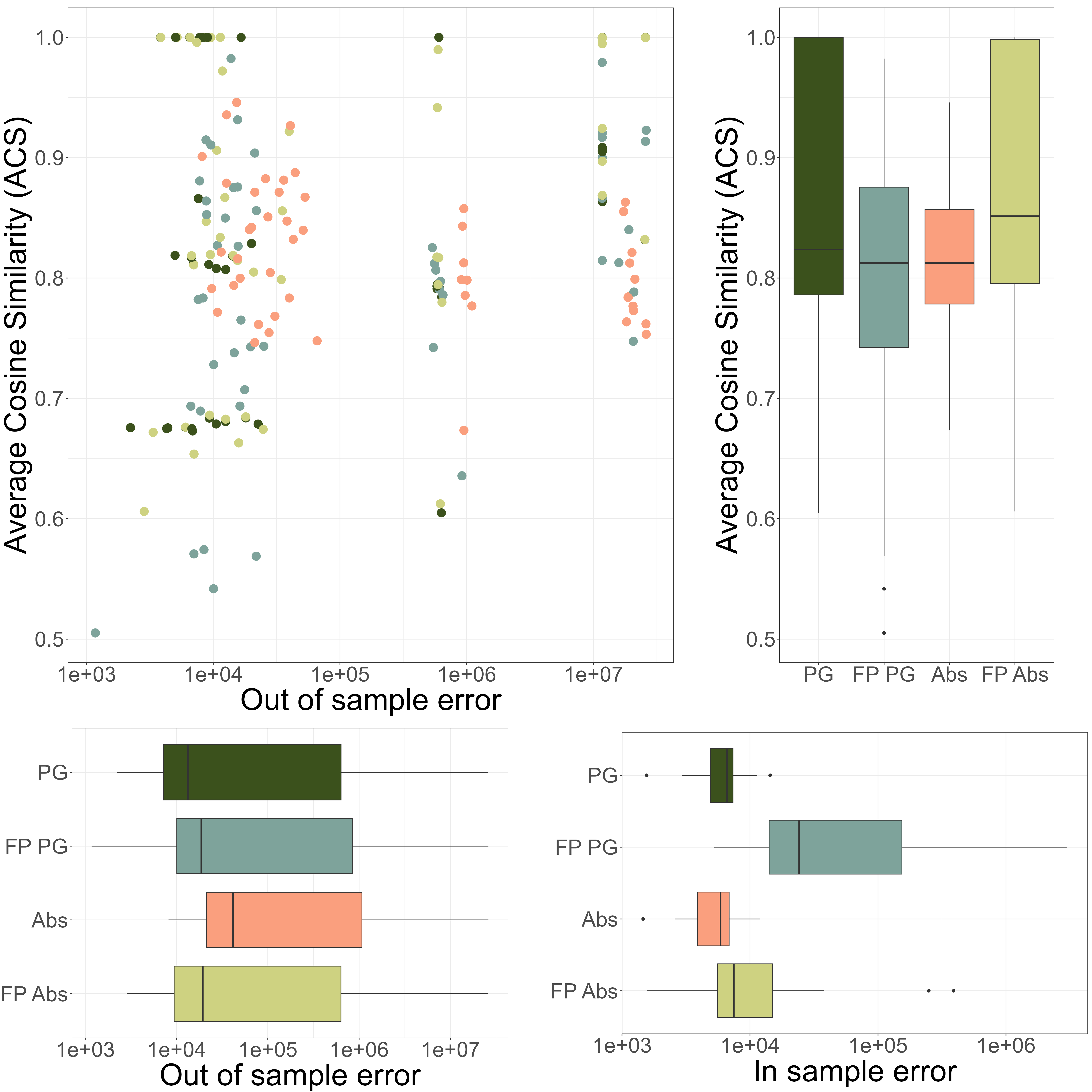}
     \caption{Top left:  Scatter plot of the average cosine similarity (ACS) between C-NMF and AE-NMF for each non-negativity constraint against the out-of-sample error of AE-NMF for each non-negativity constraint. Top right: Boxplot of the ACS between C-NMF and AE-NMF for each non-negativity constraint. Bottom plots: Boxplots of the reconstruction errors for each non-negativity constraint of AE-NMF for the test sets (left) and the training sets (right).}
     \label{fig:non_neg}
 \end{figure}

The schema enforcing non-negativity in the encoding and decoding matrices, $\bm{W}_{\text{enc}}$ and $\bm{W}_{\text{dec}}$, in AE-NMF was settled by evaluating primarily the reconstruction error and secondarily the ACS with a corresponding signature set from C-NMF. These performance measures were chosen to obtain high reconstruction accuracy and similarity with C-NMF. \newline
Dividing the 523 ovarian cancer patients' mutational profiles into 80/20 training/test set splits, 4 signatures were extracted from the training set using C-NMF and AE-NMF and each of the four non-negativity schemes (PG, FP PG, Abs, FP Abs). The training and test error for each non-negativity scheme in AE-NMF and the ACS between the C-NMF signatures and each of the AE-NMF signatures were calculated after extraction. This process was repeated for 50 train/test splits. A scatter plot of the ACS against the test errors as well as boxplots of the ACS and training and test errors for each non-negativity scheme across the 50 splits can be seen in Figure S\ref{fig:non_neg}.\newline
From these analyses it is observed that using a projected gradient (PG) or taking the absolute value in the forward pass (FP Abs) of the weight matrices perform similarly in both test error and ACS while outperforming the other methods. Methods enforcing non-negativity after each update (PG and Abs) had lower in-sample errors than methods enforcing non-negativity as a part of the forward pass (FP PG and FP Abs). Though PG appears to perform the best overall, it does experience problems by setting entire vectors to zero and being unable to exit this value again. Thus, all analyses will be carried out by taking the absolute value of negative weights in the forward pass (FP Abs) to enforce non-negativity in the weight matrices in AE-NMF.

\begin{figure}
    \centering
    \includegraphics[width = \linewidth]{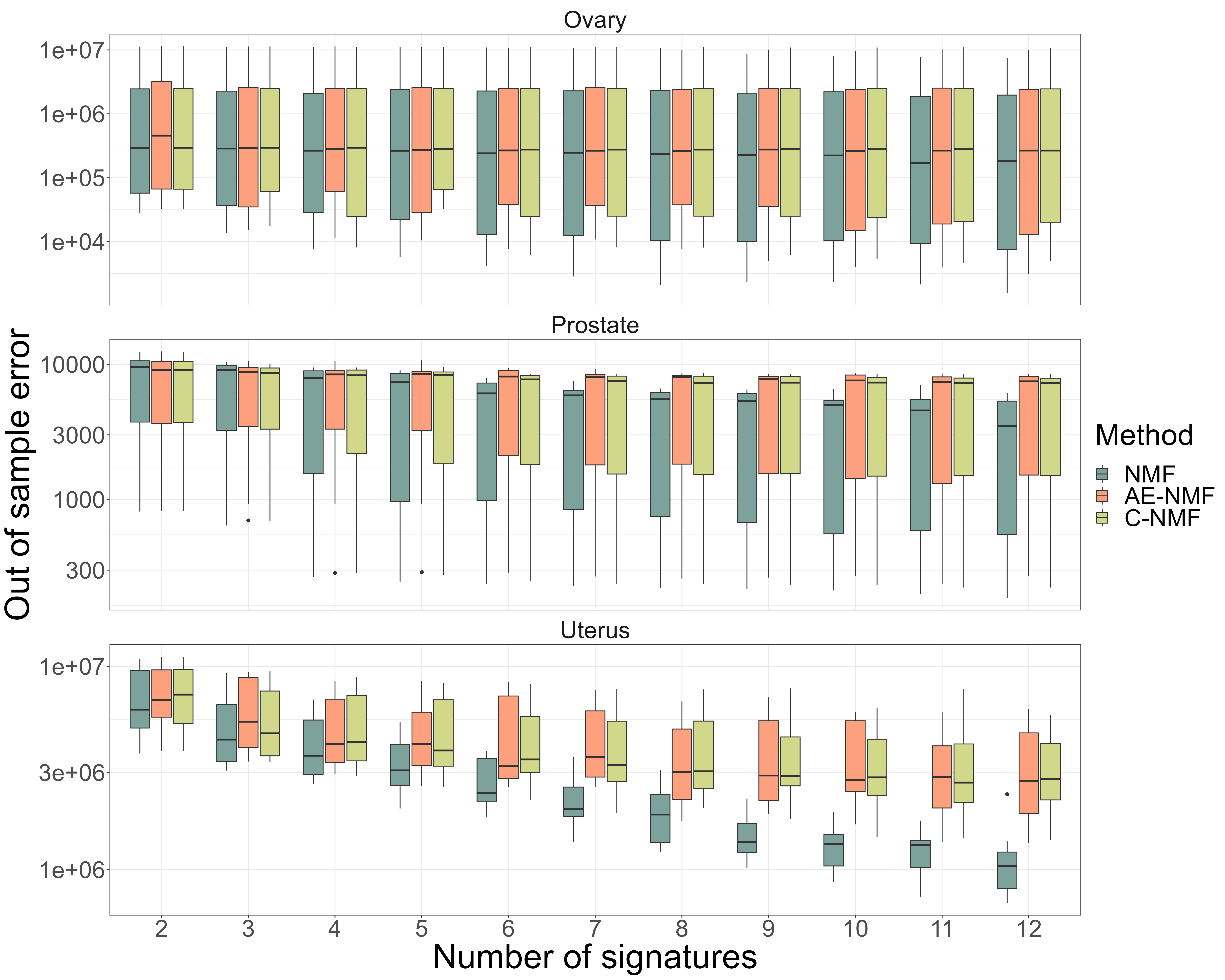}
    \caption{Boxplots of the test errors of the ten bootstrap train/test splits of the ovary, prostate and uterus cohorts against the number of signatures used in the extraction. Boxes are colored corresponding to the method used.}\label{fig:boot}
\end{figure}
 \begin{figure}
     \centering
     \includegraphics[width = 0.8\linewidth]{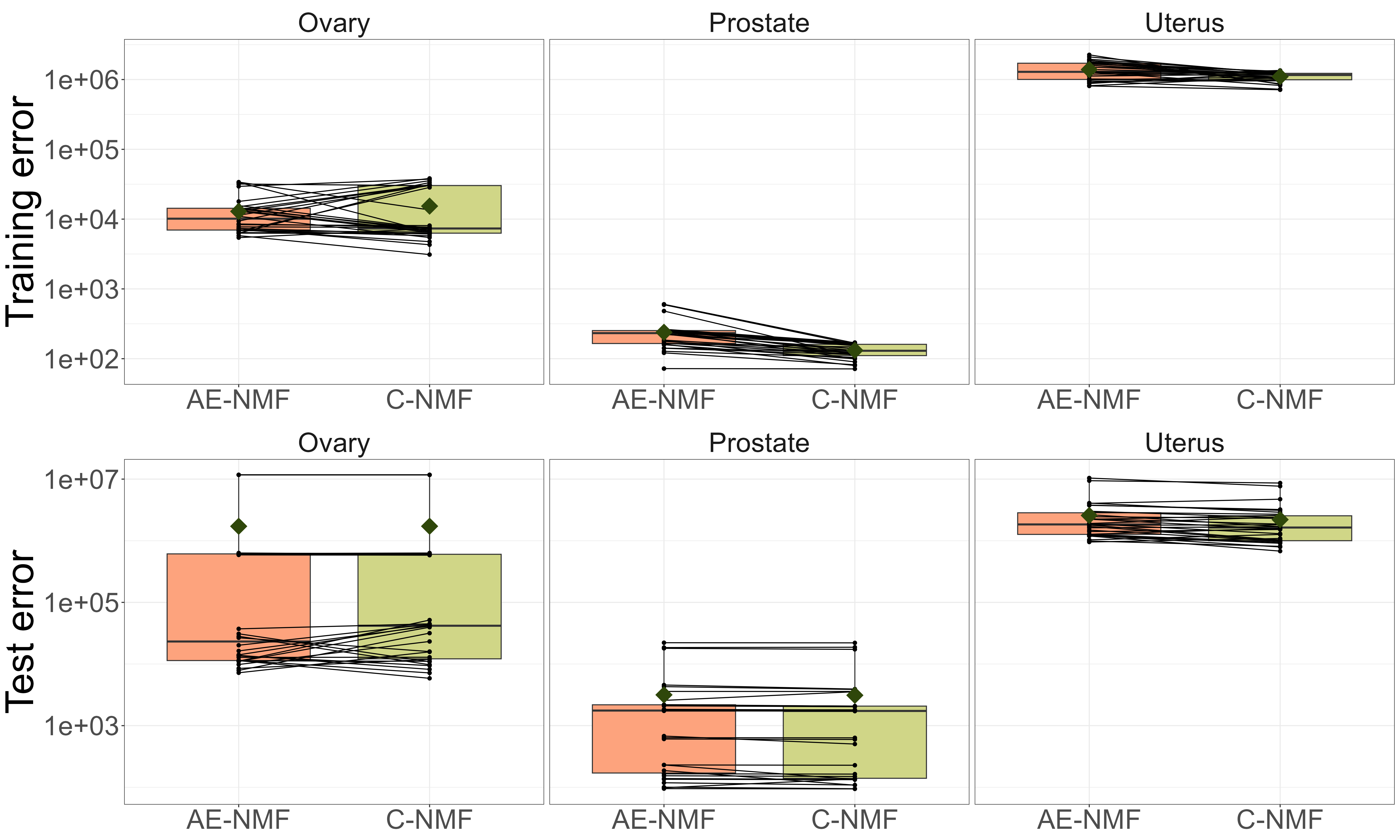}
     \caption{Boxplots of the training- and test errors of 30 train/test splits of the ovary, prostate, and uterus cohorts. C-NMF and AE-NMF errors resulting from the same splits are connected by a line. The boxes are colored corresponding to the method used and the red diamond depicts the mean error. The y-axis is on a log 10 scale.}\label{fig:error_cvx}
 \end{figure}
\begin{figure}
    \centering
    \includegraphics[width = 0.8\linewidth]{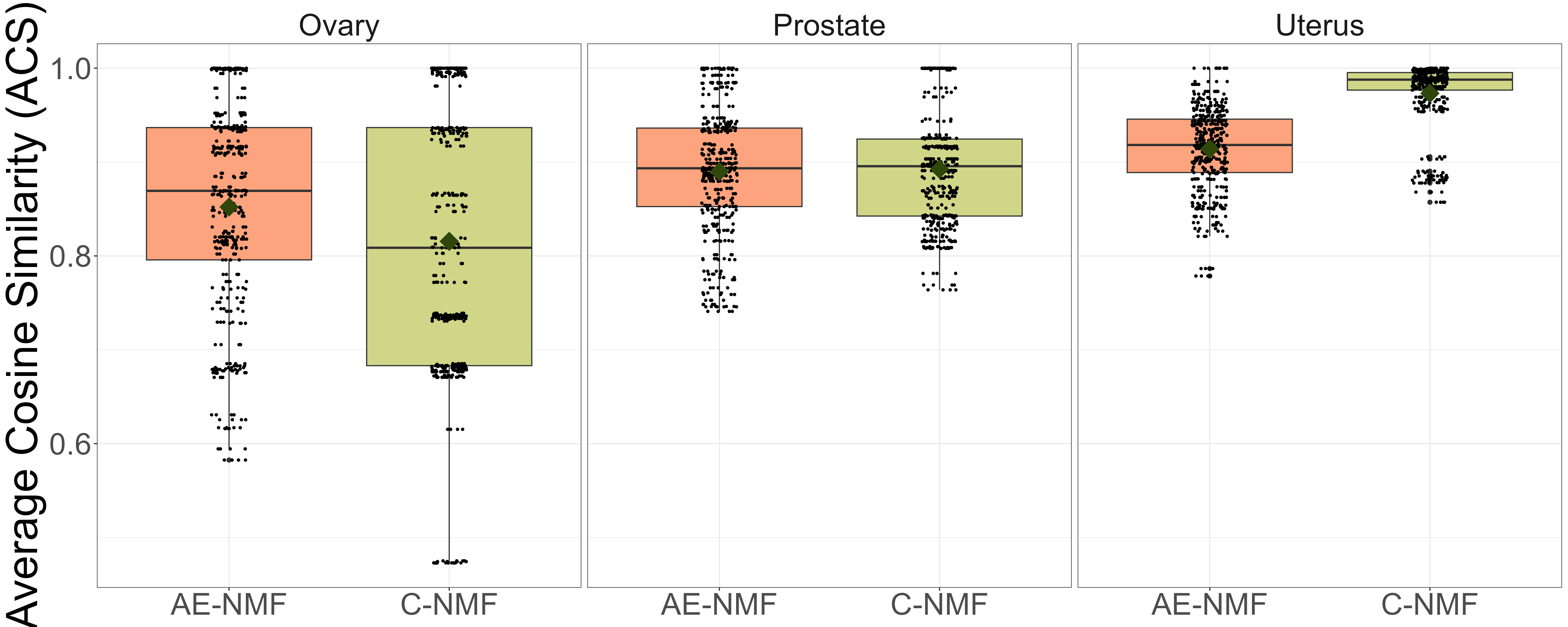}
    \caption{Box plots of the ACS between each combination of signatures extracted using a C-NMF or AE-NMF method across the 30 splits of data matrix for the ovary, prostate, and uterus cohort. Average consistency is marked by a red diamond.}\label{fig:metcos_cvx}
\end{figure}

\begin{table}[H]
\centering
\begin{tabular}{ll|rrr}
Model 1   &     & C-NMF    & C-NMF & AE-NMF \\
Model 2   &     & AE-NMF   & NMF   & NMF   \\  \hline
Data      &  K  &  \multicolumn{3}{c}{$ACS(\bm{\hat{H}}_{\text{model 1}},\bm{\hat{H}}_{\text{model 2}})$} \\ \hline
Ovary     & 4   & $\bm{0.85}$     & $0.80$  & $0.83$  \\
Prostate  & 5   & $0.90$     & $\bm{0.95}$  & $0.83$  \\
Uterus   & 8   & $\bm{0.91}$     & $0.81$  & $0.84$  \\\hline
Data      &  K  &      \multicolumn{3}{c}{$L_F(\bm{\hat{W}}_{\text{model 1}}, \bm{\hat{W}}_{\text{model 2}})$}       \\ \hline
Ovary     & 4   &  $\bm{1.1\cdot 10^{9}}$   & $\bm{1.1\cdot 10^{9}}$   & $1.2\cdot 10^9$      \\
Prostate  & 5   & $1.6\cdot 10^7$   & $1.5\cdot 10^{7}$  & $\bm{1.2\cdot 10^7}$         \\
Uterus   & 8   & $2.0\cdot 10^{10}$     & $\bm{1.4\cdot 10^{10}}$     & $1.5\cdot 10^{10}$        \\
\end{tabular}
\caption{Average cosine similarity (ACS) and Frobenius distance when pairwisely comparing the signatures and exposures, respectively, found using C-NMF, AE-NMF, and NMF on the 30 splits of all training cohorts. Highest ACS and lowest Frobenius distance are highlighted in bold.}\label{tab:comb}
\end{table}
\subsection{COSMIC Validation}
\begin{figure}
    \centering
    \includegraphics[width = \linewidth]{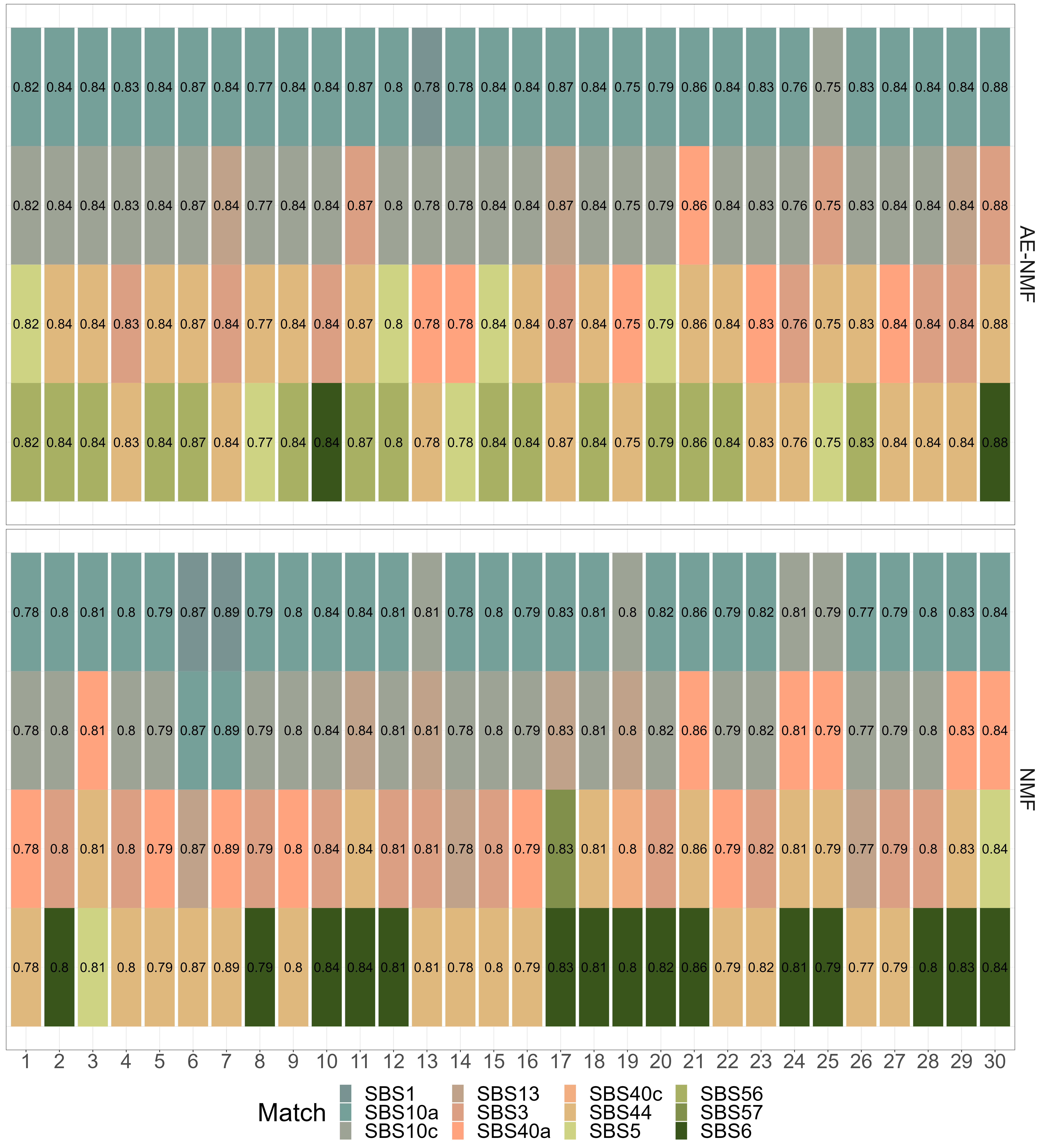}
    \caption{Cosine similarity between the four AE-NMF (top) or NMF (bottom) signatures extracted on the GEL ovary cohort and the matched COSMIC v.3.4 signature for each of the 30 train/test splits. The boxes are colored according to the matched COSMIC signature. The cosine similarity between each extracted signature and its COSMIC match is depicted within each box}\label{fig:cos_val_ov}
\end{figure}

\begin{figure}
    \centering
    \includegraphics[width = \linewidth]{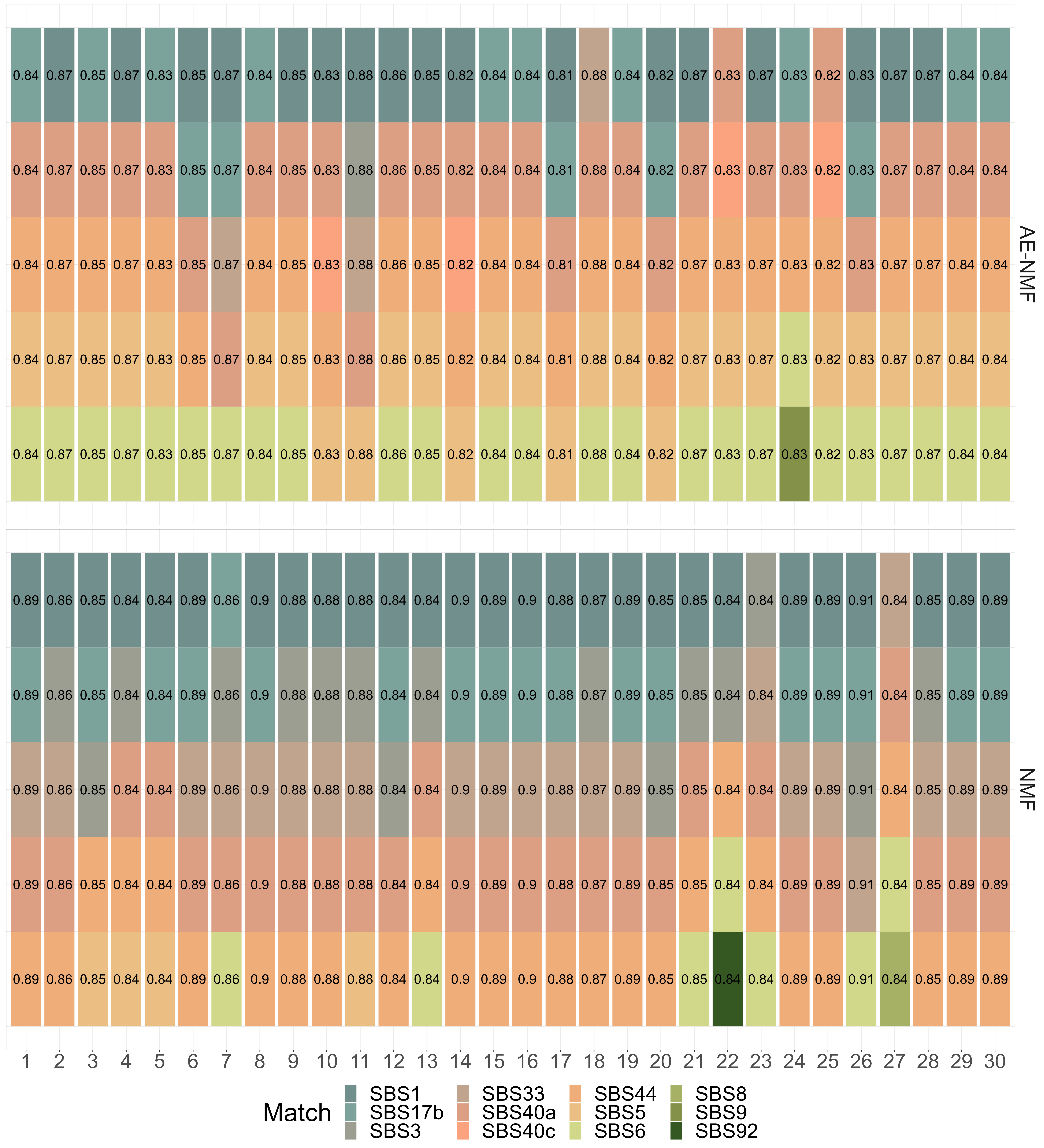}
    \caption{Cosine similarity between the 5 AE-NMF (top) or NMF (bottom) signatures extracted on the GEL ovary prostate and the matched COSMIC v.3.4 signature for each of the 30 train/test splits. The boxes are colored according to the matched COSMIC signature. The cosine similarity between each extracted signature and its COSMIC match is depicted within each box}\label{fig:cos_val_pr}
\end{figure}

\begin{figure}
    \centering
    \includegraphics[width = \linewidth]{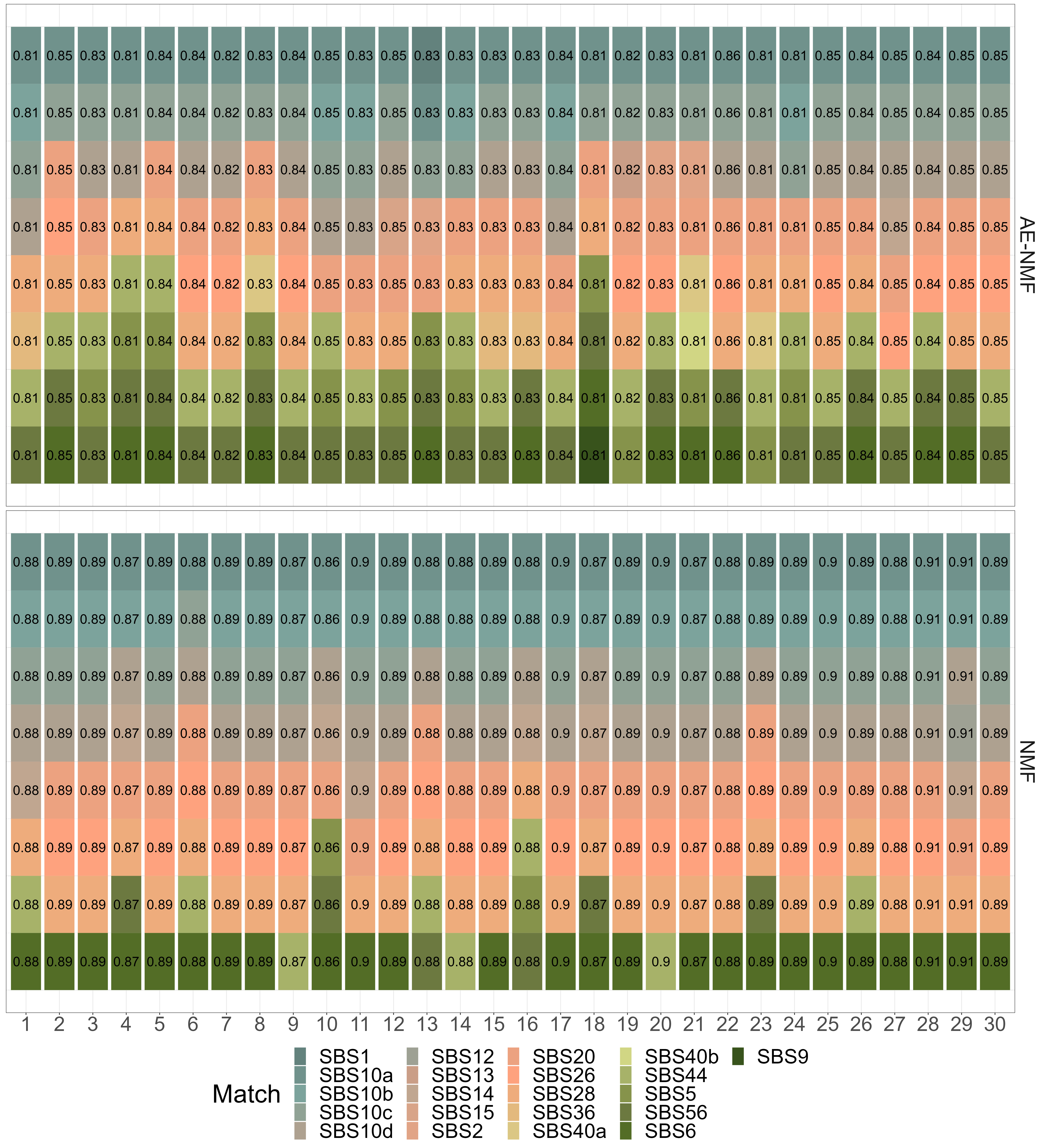}
    \caption{Cosine similarity between the 8 AE-NMF (top) or NMF (bottom) signatures extracted on the GEL uterus cohort and the matched COSMIC v.3.4 signature for each of the 30 train/test splits. The boxes are colored according to the matched COSMIC signature.  The cosine similarity between each extracted signature and its COSMIC match is depicted within each box. The cosine similarity between each extracted signature and its COSMIC match is depicted within each box}\label{fig:cos_val_ut}
\end{figure}

Figures S\ref{fig:cos_val_ov}, S\ref{fig:cos_val_pr}, and S\ref{fig:cos_val_ut} show the matched COSMIC signatures for all 30 train/test splits of the ovary, prostate, and uterus cohort, respectively. Each split is colored by the matched COSMIC signature, and the cosine similarity between the extracted signature and its COSMIC match is reported in each box.

\subsubsection{Runtime}
\begin{figure}
    \centering
    \includegraphics[width = \linewidth]{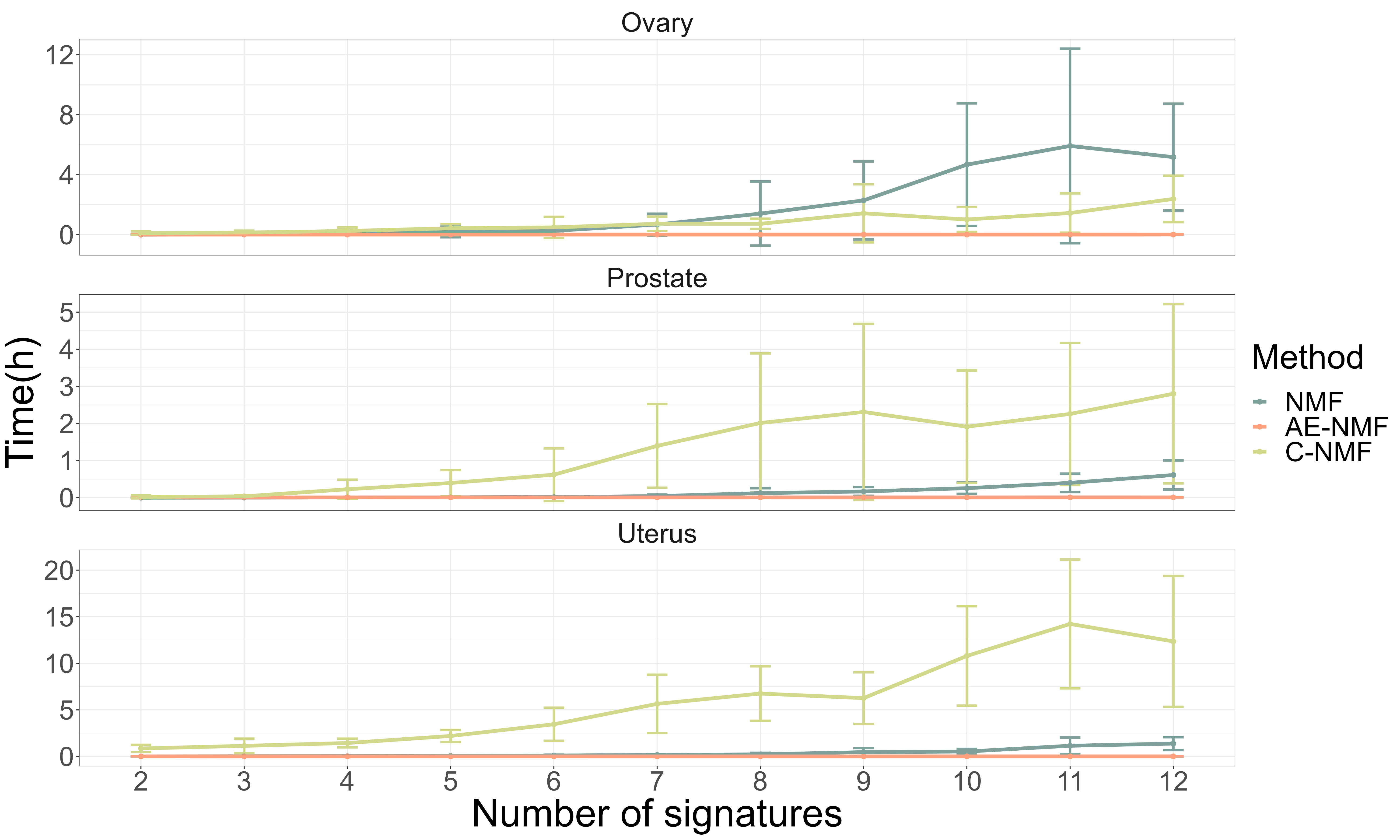}
   \caption{Mean ($\pm$ SD) running time in hours over the bootstrap samples is plotted against the number of signatures ($K$) used for the extraction of the ovary, prostate, and uterus cohorts by each method in the bootstrap analyses to determine the number of signatures.}\label{fig:timek2}
\end{figure}

While NMF and C-NMF use multiplicative updating steps in their optimization AE-NMF uses additive, gradient descent-based updating steps. In Figure S\ref{fig:timek2} the mean ($\pm$ SD) running time for signature extractions over the bootstrap samples is plotted against the number of signatures used in the extraction. We see that AE-NMF is consistently the fastest converging with the difference becoming more expressed as the number of signatures in the extraction increases. Optimizing with C-NMF becomes especially time consuming as the size of the factor matrices increases.

It it worth noting that AE-NMF updates are made with the Adam optimizer in the Python library PyTorch which is professionally constructed to be as time efficient as possible. In contrast, the code for the NMF and C-NMF updates used for this paper was written by the authors and without taking efficiency into special consideration. Therefore to be able to properly conclude on the running time of these algorithms more systematic comparisons are needed.
\FloatBarrier
\printbibliography